\documentstyle[epsf]{MN}
\def \ie        {\hbox{\it i.e.}}
\def \eg        {\hbox{\it e.g.}}
\def \cf        {\hbox{\it cf.}}

\def \etal      {et al.\ }

\def \K         { \hbox{$\,$ K} }
\def \Ho        {{\rm\ H_{0}}}
\def \kmsmpc    {{\rm\ km\ s^{-1}\ Mpc^{-1}}}
\def \kev       {{\rm\ keV}}

\def \se        {\!=\!}
\def \sims      {\sim \!}
\def \ssim      {\! \sim \!}

\def \sequiv    {\! \equiv \!}
\def \spropto   {\! \propto \!}
\def\\{\hfil\break}
\def\spose#1{\hbox to 0pt{#1\hss}}
\def\lta{\mathrel{\spose{\lower 3pt\hbox{$\mathchar"218$}}
     \raise 2.0pt\hbox{$\mathchar"13C$}}}
\def\gta{\mathrel{\spose{\lower 3pt\hbox{$\mathchar"218$}}
     \raise 2.0pt\hbox{$\mathchar"13E$}}}

\def\lesssim{\mathrel{\hbox{\rlap{\hbox{\lower4pt\hbox{$\sim$}}}\hbox{$<$}}}}
\def\gtrsim{\mathrel{\hbox{\rlap{\hbox{\lower4pt\hbox{$\sim$}}}\hbox{$>$}}}}

\newcount\itemno
\def \ino         { \the\itemno\global\advance\itemno by 1 }
\def\apj{ApJ}
\def\aj{AJ}

\def\aa{A\&A}
\def\mnras{MNRAS}

\newcommand{\beq}{\begin{equation}}
\newcommand{\eeq}{\end{equation}}
\def \xray {\hbox{X--ray} }

\def \rc {\hbox{$r_c$} }
\def \rtwoh {\hbox{$r_{200}$} }

\def \betamodel {\hbox{$\beta$--model} }

\def \logTd6 {\hbox{log$( T/6 \kev)$} }

\def \fICMbar {\hbox{$\langle f_{ICM} \rangle$}}
\def \Cb {\hbox{$C_{\rm b}$}}

\title[ Multiphase Intracluster Medium ] 
{  A Multiphase Model for the Intracluster Medium }
\author[D. Nagai, M.E. Sulkanen and A.E. Evrard]
{Daisuke Nagai$^1$, Martin E. Sulkanen$^{2,3}$, and August E. Evrard$^1$ \\
   $^1$ Physics Department, University of Michigan, Ann Arbor, MI 48109--1120 USA \\
   $^2$ X-Ray Astronomy Group, Space Sciences Laboratory, NASA/Marshall Space Flight Center,
        Huntsville, AL  35812 USA \\
   $^3$ Present address: Astronomy Department, University of Michigan, Ann Arbor, MI 48109--1090 USA \\  
daisuken@umich.edu \\
msulkanen@astro.lsa.umich.edu \\
evrard@boris.physics.umich.edu }

\date{\today}
\begin{document}

\maketitle 

\begin{abstract}

Constraints on the clustered mass density $\Omega_m$ 
of the universe derived from 
the observed population mean intracluster gas fraction $\fICMbar$
of \xray clusters may be
biased by reliance on a single-phase assumption for the thermodynamic 
structure of the intracluster medium (ICM).  We propose a descriptive
model for multiphase structure in which a spherically symmetric ICM
contains isobaric density perturbations with a radially dependent
variance $\sigma^2(r) \se \sigma_c^2 (1+r^2/r_c^2)^{-\epsilon}$.  
The model extends the work of Gunn \& Thomas (1996) which 
assumed radially independent density fluctuations thoughout
the ICM. 

Fixing the X--ray emission profile and emission weighted temperature, we
explore two independently observable signatures of the model in the 
$\{\sigma_c,\epsilon\}$ space.  For bremsstrahlung dominated emission, 
the central Sunyaev--Zeldovich (SZ) decrement 
in the multiphase case is increased over the single-phase case and 
multiphase X--ray spectra in the range $0.1-20$ keV are flatter in the 
continuum and exhibit stronger low energy emission lines than their 
single-phase counterpart.  We quantify these effects for a fiducial $10^8$ K
cluster and demonstrate how the combination of SZ and X--ray spectroscopy
can be used to identify a preferred location
$\{\hat{\sigma}_c,\hat{\epsilon}\}$ in the model plane.  From these
parameters, the correct value of $\fICMbar$ in the multiphase
model results, allowing an unbiased estimate of $\Omega_m$ to be 
recovered. 

The consistency of recent determinations of the Hubble constant from 
SZ and \xray observations with values determined by other methods 
suggests that biases in ICM gas fractions are small, $\lta 20\%$.

\end{abstract}

%taken from most recent MNRAS headings index

\section{Introduction} 

The existence of non-baryonic or ``dark'' matter on very large scales in 
the universe is inferred from a number of observations, including 
\xray and gravitational lensing observations of galaxy clusters. 
Observations suggest that the baryonic component of clusters 
predominantly consists of hot, diffuse, intergalactic medium (ICM) 
which emits X--rays by scattering of electrons in the Coulomb fields 
of electrons and ions, \ie, thermal bremsstrahlung.  
The \xray observations determine the ICM mass content 
in a model dependent fashion.  
Recent analysis of the flux limited Edge sample employs the standard,
isothermal $\beta$--model and finds a mean ICM mass
fraction $\fICMbar \se 0.212 \pm 0.006$ (Mohr, Mathiesen \& Evrard
1998; see also White \& Fabian 1996; David, Forman \& Jones 1996) 
within the virial regions of 27 nearby clusters with 
\xray temperatures above 5 keV.  This value is several 
times larger than that expected in an Einstein--deSitter universe with
the observed light element abundances (White \etal 1993).  

One way to reconcile the cluster observations with a universe having
critical mass density $\Omega_m \se 1$ is to suspect that the standard
model treatment of the ICM posseses substantial systematic errors.  
Gunn \& Thomas (1996, hereafter GT96), motivated by models of cooling flows
(Nulsen 1986; Thomas 1988) propose that a multi-phase ICM structure exists
throughout the cluster atmosphere.  A given macroscopic volume element
contains gas at a range of densities and temperatures which are assumed 
to be in pressure equilibrium.  Fixing the gas mass within this  
volume, the emission measure of a multiphase gas will increase as 
the clumping factor 
$C \sequiv \bigl<\rho^2\bigr>/\bigl<\rho\bigr>^2$.  But since we observe
luminosity, not gas mass, the implication is that clumped gas requires
less total mass $M_{gas} \spropto 1/\sqrt{C}$ in a given volume 
to produce a fixed X--ray emissivity.   

The standard analysis of the cluster plasma assumes that it
exists in a single thermodynamic phase at any location within
the cluster. In most cases an isothermal, ``beta'' model 
(Cavaliere \& Fusco-Femiano 1976) is used to describe the cluster 
plasma electron density for a spherically symmetric atmosphere.  
Under these assumptions, the observed azimuthal \xray surface 
brightness profile determines the volume emissivity at radius $r$ from
the cluster center 
\beq \label{volem}
\xi(r) \equiv \rho^2(r) \Lambda_X(T_X) = \xi_0 
\biggl ( 1 + {{r^2}\over{r_c^2}} \biggr )^{-3\beta + \frac{1}{2} }.
\eeq
Here $\xi_0$ is the central value of the \xray emissivity, $r_c$ is
the core radius of the \xray emission, $\Lambda_X (T_X)$ is the
(suitably normalized) plasma
emission function at a temperature $T_X$ over a prescribed \xray
bandwidth.  The temperature $T_X$ is determined from \xray spectral
measurements by, for example, fitting the observed spectrum to a thermal 
bremsstrahlung model.  With observations and plasma emission model 
in hand, one then constructs 
the gas mass density $\rho(r) = (\xi_0/\Lambda_X(T_X))^{1/2} 
( 1 + r^2/r_c^2)^{-3\beta/2}$ and integrates outward from the
origin to define enclosed gas mass.

The total (baryonic plus non-baryonic) 
mass within a radius $r$ is inferred from assuming that the plasma is
in hydrostatic equilibrium, supported against gravity entirely 
by thermal pressure.  The fluid equation of 
hydrostatic equilibrium then sets the total, gravitating mass 
\beq \label{hde}
M_{\rm tot} (r)  =  - {{r^2}\over{G}} {{1}\over{\rho}} {{dP}\over{dr}},
\eeq
which for the fiducial, single-phase, isothermal $\beta$--model 
cluster gives
% %
% \beq \label{mgs}
% M_{\rm tot, s} (r) = {{3}\over{2}} \beta r_c^2 {{k_B T_X}\over{\mu m_p}} 
% \biggl ( {{\xi_0 m_p^2}\over{\Lambda_X (T_X)}} \biggr )^{\frac{1}{2}} 
% {{r^2/r_c^2}\over{1 + r^2/r_c^2}}.
% \eeq
%
\beq \label{mgs}
M_{\rm tot, s} (r) = {{3\beta}\over{G}} {{k_B T_X}\over{\mu m_p}} 
{{r^3/r_c^2}\over{1 + r^2/r_c^2}}.
\eeq

In this paper, we extend a multiphase ICM 
model first proposed by Gunn \& Thomas (1996, herafter GT96) to 
incorporate radial variability in the multiphase structure.  Radial
variability is a natural expectation.  Since both cooling timescales 
($\spropto \rho^{-1}$ if nearly isothermal) and local gravitational
timescales ($\spropto \rho^{-1/2}$) increase outward from the 
cluster core, the timescale for development of multiphase structure 
should also be larger at the virial surface than in the core of a
cluster.  

We introduce the theoretical model in \S2 below.  In \S3, we examine
the effects of a multiphase structure on the mean intracluster gas
fraction $\fICMbar$ inferred from \xray observations and consider 
observable implications for the Sunyaev--Zeldovich (SZ) effect 
and \xray spectroscopy of the ICM.  For the latter, we examine two 
specific signatures --- the excess (relative to the single-phase case) 
in central SZ decrement and an X-ray spectral hardness ratio --- for the
case of a ``Coma-like'' cluster.  We show how the pairing of 
\xray spectroscopy and SZ image can be used to estimate the
magnitude of systematic error introduced into estimates of ICM 
gas fraction by assuming the standard \betamodel.  

\section{Theory}

GT96 argue that if a spectrum of plasma density fluctuations 
were generated in a cluster at a substantial fraction of a Hubble 
time in the past, then its densest phases would cool and be 
removed from the plasma.  Following Nulsen (1986), they argued that
this would produce a power-law spectrum of fluctuations 
$f(\rho) \propto \rho^{-\gamma}$ at the present, formed from a 
narrow range of phases that were initially tuned to have cooling times 
comparable to a Hubble time.  
However, this argument ignores the stochastic nature of gravitational
clustering in hierarchical models of structure formation.  In such
models, clusters grow largely by mergers of proto-cluster candidates
embedded within the large--scale filamentary network.  It is suspected
that strong mergers may, through plasma turbulence, effectively 
``reinitialize'' density fluctuations in the ICM.  Since the time
since the last major merger is a random variable in a coeval
population, then a volume limited sample will contain clusters
whose multi-phase structures are at different stages of development.   
This idea is consistent with observed properties of the local 
\xray cluster population, in which a range of central cooling 
flow behavior is present (Fabian 1994).  

Because of this and other uncertainties in the dynamical development of
multiphase structure, we postulate a log-normal form for 
the multiphase density perturbations.  We do not attempt a formal
justification for this choice;  it is motivated largely by a condition
of ``reasonableness'' and the fact that it simplifies calculations
below.  The formalism requires only low
order moments of the distribution, so the model can be recalculated
for arbirtary $f(\rho)$.  

We postulate the existence of plasma density fluctuations in a spherically
symmetric cluster atmosphere which: (i) are isobaric at a given
radius, (ii) produce a volume emission profile consistent with 
equation~(\ref{volem}) and (iii) exhibit an isothermal emission 
weighted temperature with radius.  The first item is based on a
hydrostatic assumption and the remainder impose observed 
constraints on the \xray image and emission weighted temperature 
profile.  Although isothermality extending to $r_{200}$ --- the radius
within which the mean total mass density is 200 times the critical
density --- may not be supported by observations (Markevitch \etal 1998)
or simulations (Frenk \etal 1998), temperature drops of only $10-20\%$
are allowed within $r_{200}/3$ (Irwin, Bregman \& Evrard 1998).  Since
the observables we stress in the analysis are core dominated, our
results are not particularly sensistive to departures from
isothermality which may exist near $\rtwoh$.  

\subsection{The multiphase distribution function}

We assume a log-normal form for the cluster plasma density phase 
distribution $f(\rho)\, d\rho$, the fraction of a volume element at a 
radius $r$ that contains plasma of density of between 
$\rho$ and $\rho + d\rho$, 
\beq \label{lgn}
f (\rho) \, d\rho = {{1}\over{\sqrt{2\pi} \, \sigma (r)}} \exp 
\bigg ( -{{\ln^2 [\rho/\rho_0 (r)]}\over{2 \sigma^2(r)}} 
                     \biggr )  {{d\rho}\over{\rho}}. 
\eeq
The quantity $\rho_0 (r)$ is a reference density and $\sigma^2(r)$ 
is the variance of the distribution.  Since the core radius presents a
characteristic scale in the \xray image, we take a form 
\beq \label{sgr}
\sigma^2 (r) = \sigma_c^2 (1 + r^2/r_c^2 )^{-\epsilon},
\eeq
for the variance, with $r_c$ the core radius of the beta-model 
density profile described earlier, 
and $\sigma_c$ and $\epsilon$ are free parameters which  
set the magnitude and radial dependence of the multiphase structure.  
We consider 
such a parameterization in order to couple the magnitude of density 
fluctuations to the likelihood that the local conditions have allowed 
cooling to amplify them.  A simple parameterization is one in which 
the variance scales with the inverse of the local cooling time 
$\sigma^2(r) \spropto \tau_{\rm cool}^{-1}(r)$.  An isothermal atmosphere 
(for which $\tau_{\rm cool}(r) \spropto \rho^{-1}(r)$) will have 
$\sigma^2(r) \spropto \rho(r)$, implying
$\epsilon \se 3\beta/2 \se 1$ for the characteristic $\beta \se
2/3$ value seen in \xray images.  We consider values in the range
$\epsilon \in 0-1$.  The limit $\epsilon \rightarrow \infty$
represents a multiphase structure existing purely within the cluster
core.  
In the limit $\sigma_c \rightarrow 0$, we recover a 
single-phase plasma for any value of $\epsilon$, 
while the limit $\epsilon \rightarrow 0$ yields the
multiphase model results (no position dependence) of GT96.

The definition of $\rho_0(r)$ is now absorbed into the specification of
the mean density at radius $r$
\beq \label{rom}
\langle \rho(r) \rangle \equiv \int {\rho \, f(\rho) \, d\rho} =
                       \rho_0 \exp \biggl ( {{1}\over{2}} \sigma^2 (r) \biggr ),
\eeq
where $\langle \rangle$ represents an ensemble average of volume elements
on a spherical shell of radius $r$. A useful equation is a generalization 
of equation (\ref{rom}) to higher moments, namely
\beq \label{ron}
\langle \rho^q \rangle = \int {\rho^q \, f(\rho) \, d\rho} =
                       \langle \rho \rangle^q \, \exp \biggl ( {{q(q-1)}\over{2}} 
\sigma^2 (r) \biggr ).
\eeq

\subsection{A multiphase ``isothermal \betamodel'' cluster}

We now impose some observational constraints on the model.  
Assuming a power--law emissivity function 
\beq \label{LambdaX}
\Lambda_X(T) \propto T^{\alpha} ,
\eeq
the requirement that the emission weighted temperature profile be
isothermal at temperature $T_X$ implies that the condition 
\beq \label{txdef}
T_X \equiv { {\langle \rho^2 T^{1+\alpha} \rangle} \over
            {\langle \rho^2 T^\alpha \rangle} }
\eeq
holds at all cluster radii. 

Under an ideal gas assumption $P = (\rho/\mu m_p) k_B T$, with $m_p$
the proton mass and $\mu$ the mean molecular weight, 
equations (\ref{txdef}) and (\ref{ron}) can be used to define the 
local gas pressure in the multiphase medium 
\beq \label{pdef}
P(r) = {{k_B T_X}\over{\mu m_p}} \langle \rho(r) \rangle \, \exp [(1-\alpha)\sigma^2(r)].
\eeq

We now equate the known emission profile of the cluster,
equation~(\ref{volem}), to the ensemble-averaged value of the emissivity 
\beq \label{ems}
\xi_0 \bigl ( 1 + {{r^2}\over{r_c^2}} \bigr )^{-3\beta} =
{{\Lambda_0}\over{m_p^2}} \bigl ( {{\mu m_p}\over{k_B}} \bigr )^\alpha \langle 
\rho(r)^{2-\alpha} \rangle P^\alpha (r).  
\eeq
The combination of equations~(\ref{txdef}) and (\ref{ems}) 
is the canonical ``isothermal \betamodel'' assumption.  
From an observer's perspective, a 
multiphase cluster in these two measures is indistinguishable from the 
single-phase case.  Equation~(\ref{ems}) can be rearranged to give
\begin{eqnarray} \label{rnd}
\langle \rho(r) \rangle  \!\!\!\! & = & \!\!\!\! \biggl( {{\xi_0 
                           m_p^2}\over{\Lambda_X(T_X)}} \biggr)^{1/2}
                            \biggl ( 1 + {{r^2}\over{r_c^2}}
\biggr )^{-\frac{3}{2}\beta} \nonumber \\ 
&  &\exp \bigl ( {{(\alpha-1)(\alpha+2)}\over{4}} \sigma^2 (r) \bigr).  
\end{eqnarray}
Although this now defines the characteristic density $\rho_0$ used in
equation~(\ref{lgn}), it is better to identify the limit $\sigma^2(r)
\rightarrow 0$ as the single-phase density.  Following GT96, we 
introduce a multiphase ``correction factor'' for the gas mass $C_\rho(r)$
which relates the mean gas density in the multiphase case $\rho_m (r)$ 
to its single-phase value $\rho_s (r)$ 

\beq \label{rhomp}
\rho_m (r)  \equiv  \langle \rho (r) \rangle \equiv C_\rho
(r) \rho_s (r) .
\eeq
Equation~(\ref{rnd}) then implies
\beq \label{crdef}
C_\rho (r) = \exp \bigl ( {{(\alpha-1)(\alpha+2)}\over{4}} \sigma^2 (r) \bigr ) .
\eeq

A similar exercise for the gas pressure 
\beq \label{Pmp}
P (r) \equiv  C_P (r) \bigl ( {{k_B T_X}\over{\mu m_p}} \bigr ) \rho_s (r) 
\eeq
yields 
\beq \label{cpdef}
C_P (r) = \exp  \bigl ( {{(1-\alpha)(2-\alpha)}\over{4}} \sigma^2 (r) \bigr ).
\eeq
Note that for values of the \xray emission exponent $\alpha < 1$, 
the multiphase gas mass 
is lower than that of the single-phase model while the multiphase pressure
is greater than the single-phase pressure. This arises because the high-density
phases are more efficient in producing a given \xray power (provided the
emission is only a weak function of temperature).  Since the emission
weighted $T_X$ reflects the temperature in higher than average density 
regions, the pressure at all radii is increased over the single-phase case.  

The cluster gas mass for the multiphase model within a radius $r$ is given by 
\beq \label{mgdef}
M_{\rm gas, m} (r) = 4\pi \int_0^r C_\rho (r^\prime) \, \rho_s
(r^\prime) \, {r^\prime}^2 d r^\prime, 
\eeq
so that the enclosed gas mass for the multiphase model differs from 
the single-phase case by the factor
\begin{eqnarray} \label{cgdef}
C_{\rm gas} (r) & \equiv & {{M_{\rm gas, m}(r)}\over{M_{\rm gas, s}(r)}} \nonumber \\
& = & {{\int_0^{r} C_\rho (r^\prime) \, \rho_s (r^\prime) \, {r^\prime}^2 dr^\prime}\over
{\int_0^{r} \rho_s (r^\prime) \, {r^\prime}^2 dr^\prime}}.
\end{eqnarray}

The total mass of the cluster within a radius $r$ is determined from the
hydrostatic equilibrium, equation~(\ref{hde}), when comibined with equations 
(\ref{rnd})-(\ref{cpdef}) give
\begin{eqnarray} \label{mtdef}
M_{\rm tot, m} (r) & = & - \biggl ( {{r^2}\over{G}} \biggr ) 
\biggl ( {{C_P (r) P_s' (r) + P_s (r) C_P' (r)}\over{C_\rho (r) \rho_s (r)}} \biggr ) 
\nonumber \\
& = & {{C_P (r)}\over{C_\rho (r)}} M_{\rm tot,s} (r) + {{r^2}\over{G}} 
{{k_B T_X}\over{\mu m_p}}
\biggl | {{C_P' (r)}\over{C_\rho (r)}} \biggr |.
\end{eqnarray}
The total cluster mass for the multiphase model differs 
from that of the single-phase model by the factor
\beq \label{ctdef}
C_{\rm tot} (r) = {{C_P (r)}\over{C_\rho (r)}} + {{r^2}\over{G M_{\rm tot,s} (r)}}
{{k_B T_X}\over{\mu m_p}}
\biggl | {{C_P' (r)}\over{C_\rho (r)}} \biggr |.
\eeq
For bremsstrahlung emission ($\alpha \simeq 0.5$), the gas mass is 
decreased and the total mass increased in the multiphase case,
implying the enclosed gas fraction at radius $r$ is lower than 
that for a single-phase medium by the factor 
$C_{\rm b}(r) = C_{\rm gas}(r) / C_{\rm tot}(r)$.

\begin{figure}
\vskip 0.0 truecm
\epsfxsize=8.0cm
\epsfysize=8.0cm
\hbox{\hskip 0.5 truecm \epsfbox{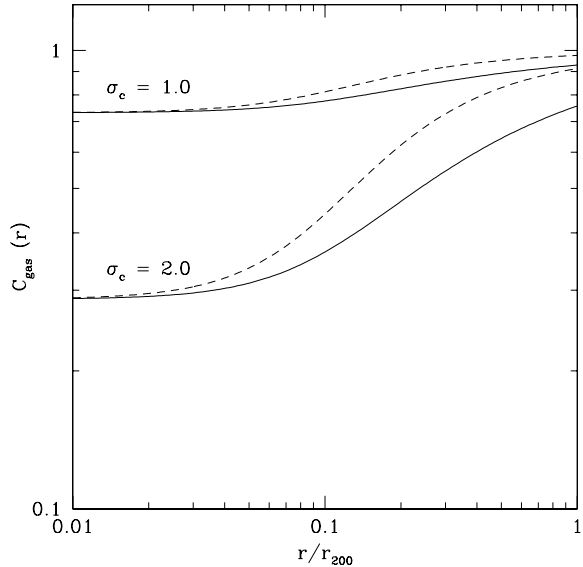} }
\vskip -0.2 truecm
\caption{
The gas mass multiphase correction factor $C_{gas}(r)$,
equation~(\ref{cgdef}), is plotted as a
function of the dimensionless radius $r/r_{200}$ assuming parameters
$r_c/r_{200}=0.1$, $\alpha=1/2$ and $\beta=2/3$.  
The solid and dotted lines represent $\varepsilon=\frac{1}{2}$ 
and $\varepsilon=1.0$, respectively, and values of $\sigma_c$ are as 
indicated.
}
\label{gasmass}
\end{figure}

\begin{figure}
\vskip 0.0 truecm
\epsfxsize=8.0cm
\epsfysize=8.0cm
\hbox{\hskip 0.5 truecm \epsfbox{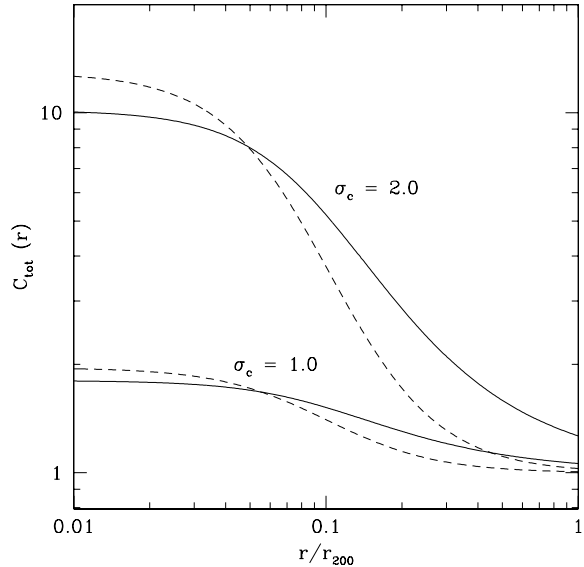} }
\vskip -0.2 truecm
\caption{
The total mass multiphase correction factor $C_{tot}(r)$,
equation~(\ref{ctdef}) is shown as a function of radius, using the
same parameters and format as in Figure~\ref{gasmass}.  
}
\label{totmass}
\end{figure}

Figure~\ref{gasmass} plots the correction factor for the cluster gas
mass, $C_{\rm gas}$, for a few choices of the controlling parameters 
$\sigma_c$ and $\epsilon$.  For purposes of illustration, we take 
structural parameters representative of rich clusters, 
namely $r_c/r_{200} \se 0.1$ and $\beta \se 2/3$ (Neumann \& Arnaud
1999), and assume pure bremsstrahlung emission, $\alpha \se 1/2$.   
The effect of the radial falloff of the multiphase structure on gas
mass estimates is substantial.  Density variations with large central
rms perturbations $\sigma_c \simeq 2.0$ produce substantial (factor
$\sims 3$) relative correction to the gas mass in the cluster core, 
but the effect on the total virial gas mass (mass within \rtwoh) 
is reduced to $25\%$ if 
$\epsilon \se \frac{1}{2}$ and only $10\%$ if $\epsilon \se 1$.   
Degeneracies exist in the virial gas correction factor; 
a relatively weak multiphase plasma distributed throughout the cluster
can produce an effect that is similar to a plasma with strong density
variations concentrated toward the center of the cluster ({\sl cf.\/} 
$\{\sigma_c, \epsilon\}$ combinations of $\{1,\frac{1}{2}\}$ and
$\{2,1\}$).  

The correction factor for the total cluster mass, 
$C_{\rm tot}$, for the same multiphase parameters is shown in
Figure~\ref{totmass}.  By steepening pressure gradients, 
the multiphase effects increase the total cluster mass 
derived from equation (\ref{mtdef}).
Once again, weaker multiphase effects distributed throughout the 
cluster can yield a total mass within $r_{200}$ that is similar 
to a cluster plasma with stronger density variations concentrated 
in the cluster center. However, such concentrated multiphase effects 
will produce a total mass profile that is steeper (\cf $(\sigma_c,\epsilon)$
of $(2,\frac{1}{2})$ {\sl vs.} $(2,1)$).   Observations of 
strong gravitational lensing could be used to break this parameter 
degeneracy, to the extent that the hydrostatic
assumption is valid in the cluster core.

% \subsection{Modeling \xray spectra}

% \subsection{Sunyaev-Zel'dovich Effect}

\section{Consequences}

We now turn to the issue of the effect of multiphase structure on
inferred ICM gas fractions and cluster observables.  
For the latter, we consider the effects of multiphase plasma on a 
cluster's \xray spectrum and the Sunvaev-Zel'dovich microwave 
decrement through a line-of-sight taken through the center of the cluster.

All of the results we discuss assume a standard structure model with 
core radius for the broadband \xray emissivity (equation (\ref{volem})) of 
$r_c = 0.1 r_{200}$, exponent $\beta = 2/3$, and, for creation of
\xray spectra, an emission-weighted \xray temperature $T_X = 10^8$ K.
Unless otherwise stated, we employ a value of $\alpha \se 0.36$ for 
the exponent of the plasma emission function, derived from a
Raymond-Smith code as described below.  Since we ignore galaxies in
our modeling, the ICM gas fraction is synonymous with the cluster 
baryon fraction.  We use the terms interchageably below, but it must
be remembered that the stellar content of cluster galaxies and
intracluster light presents an absolute lower limit to the baryon
content of clusters.

\subsection{Baryon fraction bias}

\begin{figure}
\vskip 0.0 truecm
\epsfxsize=8.0cm
\epsfysize=8.0cm
\hbox{\hskip 0.5 truecm \epsfbox{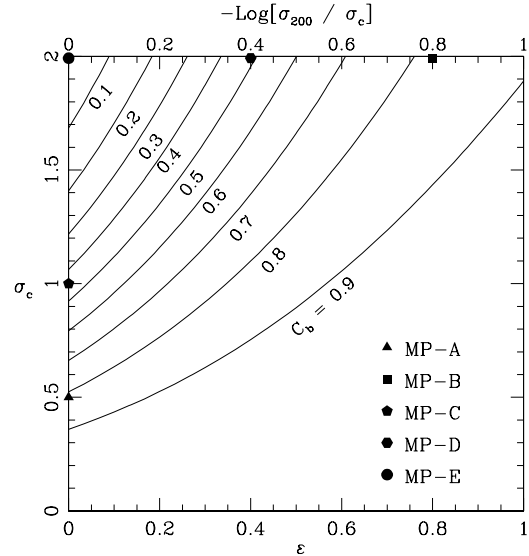} }
\vskip -0.3 truecm
\caption{The baryon multiphase correction factor evaluated at the
virial radius, $C_b (r_{200})$ is displayed in a contour plot within 
the $\varepsilon$, $\sigma_c$ plane. 
Standard parameters $\rc/r_{200} \se 0.1$ and $\beta \se 2/3$ 
are assumed.  Labels refer to models whose multi-phase spectra are
displayed in Figures~\ref{xrayspec} and \ref{rsband}.  
}
\label{cbarycont}
\end{figure}

The effects of increased total mass and decreased gas mass shown in
Figures~\ref{gasmass} and \ref{totmass} combine multiplicatively 
to reduce the cluster baryon fraction.   The magnitude of the 
effect within the virial radius is shown in Figure~\ref{cbarycont},
where we show contours of $C_{\rm b} (r_{200})$, the baryon reduction 
factor, in the $ \{ \sigma_c, \epsilon \}$ plane.  

The baryon reduction effect peaks at high $\sigma_c$ and low
$\epsilon$.   At $\epsilon \se 0$, the uniform, mulitphase structure of
GT96 is recovered, with magnitude
\beq \label{Cbeps0}
C_{\rm b} = \frac{C_{\rho}^2}{C_{\rm P}} = 
 \exp  \bigl ( {{(\alpha-1)(\alpha+6)}\over{4}} \sigma_c^2 \bigr ).
\eeq
To reduce the baryon fraction by factors $\Cb \gta 2$ requires 
density variations of magnitude $\sigma_c \gta 1$.   

In the following discussion, we highlight 
a set of five specific models, listed in Table~1.  Models A and B 
have small ($\sims 20\%$) baryon corrections, models C and D have
large baryon bias $\Cb \ssim 2$ and model E is an extreme model in
which the baryon fraction is reduced by an order of magnitude.

\begin{table} 
\caption{Model Definitions \label{fgastable}}
\begin{flushleft} 
\begin{tabular}{lcccc} 
Label  & $\sigma_c$ & $\epsilon$ \\
\hline
SP & $ 0 $ & $ - $\\
MP-A  & $ 0.5 $ & $ 0 $ \\
MP-B  & $ 2.0 $ & $ 0.8 $ \\
MP-C  & $ 1.0$ & $ 0 $ \\
MP-D  & $ 2.0 $ & $ 0.4 $ \\
MP-E  & $ 2.0$ & $ 0 $ \\
\end{tabular}
\end{flushleft}  
\end{table}

\subsection{Sunyaev-Zel'dovich Effect}

The SZ effect is produced by inverse-Compton 
scattering of cosmic microwave background (CMB) radiation off thermally
excited electrons in the hot ICM plasma (see Birkinshaw 1998 for a
recent review).  We calculate the central Comptonization parameter of 
the (nonrelativistic) thermal SZ effect 
\beq \label{ydef}
y(0) = \int \, n_e(l) \sigma_T {{k_B T(l)}\over{m_e c^2}} \, dl
\eeq
where the integral $dl$ is along a narrow line of sight through the center 
of the spherical cluster.  Here $n_e \se \rho/\mu_e m_p$ is the electron number
density and $\sigma_T$ the Thomson cross section.  Since the plasma phases
are assumed isobaric, the product $n_e(r) T(r)$ is constant, and no
phase integral is necessary in the multiphase case.  
Deviation of the $y$--decrement from that of a single-phase
plasma is caused by the alteration of the overall pressure profile in 
the cluster.  

\begin{figure}
\vskip 0.0 truecm
\epsfxsize=8.0cm
\epsfysize=8.0cm
\hbox{\hskip 0.5 truecm \epsfbox{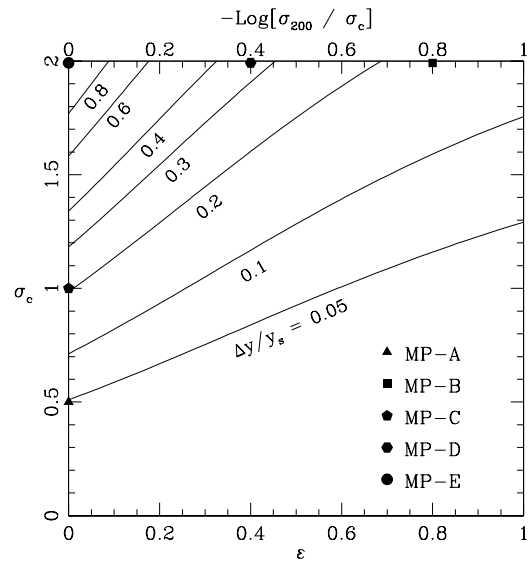} } 
\vskip -0.3 truecm
\caption{Contours of the fractional increase in central
Sunyaev-Zel'dovich decrement for the multiphase models relative to the
singe-phase case, assuming $\rc/r_{200} \se 0.1$,
$\beta  \se 2/3$ and $\alpha \se 0.36$.  }
\label{SZcontor}
\end{figure}

The fractional deviation of the central Comptonization parameter 
$\Delta y/y_s \se [y_m(0) - y_s(0)] / y_s(0)$ 
in the multiphase with respect to the single-phase model 
is shown in Figure~\ref{SZcontor}.  
In the case of a uniform, multiphase structure ($\epsilon \se 0$), the
fractional change in $y$ follows from the pressure correction factor,
equation~(\ref{cpdef}), 
\beq \label{deltay_eps0}
\Delta y/y_s  = 
 \exp  \bigl ( {{(1-\alpha)(2-\alpha)}\over{4}} \sigma_c^2 \bigr ) - 1 .
\eeq
Figure~\ref{SZcontor} shows data for the case $\alpha \se 0.36$,
approximately the slope of the $2-10$ keV luminosity versus
temperature, derived from a Raymond--Smith plasma code assuming
a one--third solar abundance of metals.  Similar values result for the
case $\alpha \se 0.5$.  

The two models with $20\%$ baryon diminution have modest, but
potentially discernable, SZ effects.  Model A has a central decrement 
enhanced by $10\%$ while model B is enhanced by $25\%$ over the
single-phase case.  The latter is similar to the $30\%$ effect for 
model C, one of the large baryon fraction diminution models.  
The other factor 2 baryon fraction model --- model D --- has a central
value of $y$ increased by $\sims 50\%$ over the standard \betamodel.  
The extreme model E has a signal enlarged by a factor 2 over the
standard case.  

Even in the single-phase case, there is inherent uncertainty in
predicting the SZ effect amplitude from X-ray observations which
arises from uncertainty in the physical distance to the cluster.  At
low redshifts, the distance error is completely due to uncertainty 
in the Hubble constant.  Given a cluster with fixed \xray properties,
a fractional deviation in central SZ decrement $\Delta y/y_s$ due to a
multiphase medium could, instead, be interpreted as a distance
effect.  This would imply a fractional error in the Hubble constant 
\beq \label{H0}
\Delta H_0/H_0 \ = \ -2 \ \Delta y/y_s  .
\eeq
For example, in a universe with true Hubble constant of $65 \kmsmpc$,
observations of a multiphase model A cluster would yield a value 
$52 \kmsmpc$ and model B would produce an estimate of $42$.  The other
highlighted models would produce even lower estimates of $\Ho$.  

Note that this result has the {\sl opposite} sense of correction 
for the SZ decrement compared to other estimates of the SZ effect 
with multiphase gas (\eg. Holzapfel \etal 1997). This is because the 
gas pressure for isobaric density fluctuations is greater than that 
for a single-phase medium, whereas other models with adiabatic 
density fluctuations, such as those present in SPH calculations
without cooling (Inagaki, Suginohara \& Suto 1995),  
have a pressure lower than that of the single-phase gas.

\subsection{X-ray spectra}

\begin{figure}
\vskip 0.0 truecm
\epsfxsize=8.0cm
\epsfysize=8.0cm
\hbox{\hskip 0.5 truecm \epsfbox{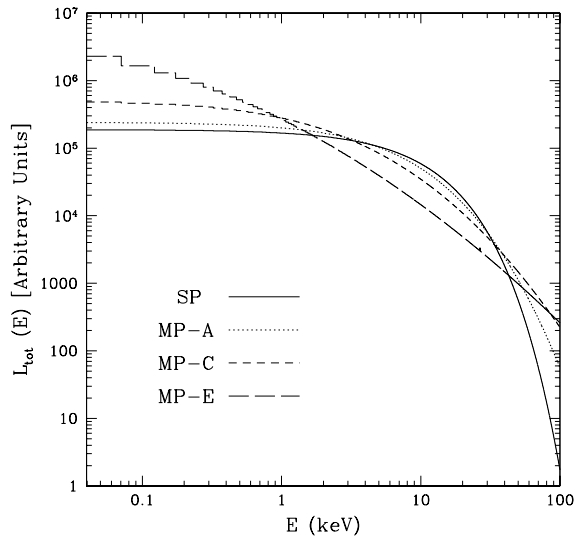} }
\vskip -0.2 truecm
\caption{X-ray continuum emission for different multiphase 
models listed in Table 1.  A value of $\alpha \se 0.5$ is assumed.
All models have the same emission weighted temperature of $10^8 \K$ in
the spectral region shown. 
}
\label{xrayspec}
\end{figure}

Spectroscopic analysis of the \xray emission provides an alternative, 
independent diagnostic of multiphase structure.  
We calculate expected \xray spectra from the multiphase plasma in two
ways.  We first consider a simple model for the 
\xray bremsstrahlung continuum from the cluster, using an 
emission function of a purely hydrogen plasma, with 
$\varepsilon (E,T) \propto T^{-\frac{1}{2}} e^{-\frac{E}{kT}}$ 
and a Gaunt factor of unity.  The plasma emission function is 
then $\Lambda (T) \se K_0 T^{1/2}$, with $K_0$ an arbitrary
normalization amplitude.  Second, 
a more detailed spectrum is calculated, 
using the Raymond-Smith plasma emission code (Raymond and Smith 1977) with 
metal abundances one-third of the solar value (Allen 1973).  The
former approach highlights continuum behavior while the latter 
allows the study of the behavior of \xray lines between $0.5 - 9$ keV.

We generate the composite spectrum emitted by the multiphase 
cluster atmosphere by numerically integrating the weighted emission over 
the density distribution at a particular radius, 
then intergrating over the cluster volume
$V = \frac{4\pi}{3} \, r^3_{200}$.   Gas beyond the virial radius
$\rtwoh$ is ignored.

\begin{figure}
\vskip 0.0 truecm
\epsfxsize=8.0cm
\epsfysize=8.0cm
\hbox{\hskip 0.5 truecm \epsfbox{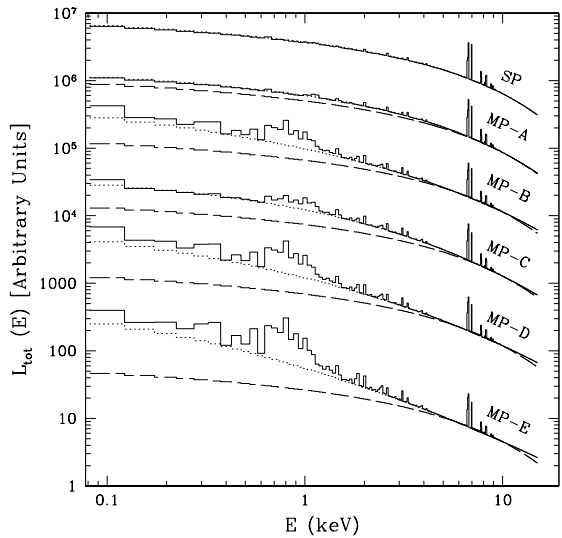} }
\vskip -0.2 truecm
\caption{X-ray spectra for different multiphase models derived from 
a Raymond-Smith code.   All models have
the same emission weighted temperature of $10^8 \K$ in
the $2-10 \kev$ region. The luminosity of each model is 
displaced vertically by arbitrary amounts for clarity.  
Solid lines assume a one--third solar abundance
plasma while dotted lines show the multiphase emission assuming zero
metallicity.  For reference, long dashed lines show the \xray spectrum
of the single-phase cluster with no metal abundances. }
\label{rsband}
\end{figure}

The appearance of the continuum plasma \xray emission in the
multiphase case can differ substantially from that of the isothermal, 
single-phase plasma.  In Figure~\ref{xrayspec}, we show the $0.05-100$ keV
bolometric \xray spectra of three multiphase models (A, C and E) 
along with the single-phase case.  All spectra are
normalized to yield the same emission weighted temperature of 
$10^8 \K$.   We assume all phases are optically thin.  
The most important effect on the spectra is the 
appearance of both low-energy ($E \ll k_B T_X$)
and high-energy ($E \gg k_B T_X$) enhancements of the spectrum with 
increasing magnitude of multiphase effects.  This shape change arises
from the blending of gas at temperatures both below and above the
fiducial $10^8 \K$ value.  
In the limit of extreme multiphase strength (model E), the
bremsstrahlung spectrum approaches power-law like behavior. 

A more complete X-ray spectrum of clusters is code.  In particular, the use of such a 
code allows invesigation of line emission as a diagnostic of
multiphase structure.  

Figure~\ref{rsband} shows the simulated emission, derived from 
a Raymond-Smith code, between
$0.1-15$ keV photon energies for a plasma with an assumed metallicity
equal to one--third of solar abundance.  Along with the rise of the
low-energy continuum, the other prominent effect  
of increased multiphase structure is the strengthening of low-energy 
($\sims 1$ keV) emission lines.  
To highlight line versus continuum effects, we plot both 
zero and one--third solar metallicity predictions for the emission for 
each model shown.  

The complex of lines between $0.5$ and $1.5$ keV presents a useful
diagnostic for multiphase structure.  
Included in this region of the spectra are 
the Fe L-shell lines, as well as H-like and He-like emission from N, O,
Ne and Mg.  For example, weak baryon bias 
models (A and B) are readily distinguished by this emission signature,
as are models the more strongly multiphase models, C and D.  

In contrast to the low energy lines, the strength of the 
7 keV iron complex is almost unaffected by multiphase structure.
These lines originate in hot phases very close to the fiducial
temperature of $10^8 \K$.  The emission weighted
temperature constraint imposed on the models requires that the 
contribution to the total emission from phases near the fiducial
temperature cannot vary by large factors.  Hence the hot emission
lines do not vary significantly among the multiphase models.  

\begin{figure}
\vskip 0.0 truecm
\epsfxsize=8.0cm
\epsfysize=8.0cm
\hbox{\hskip 0.5 truecm \epsfbox{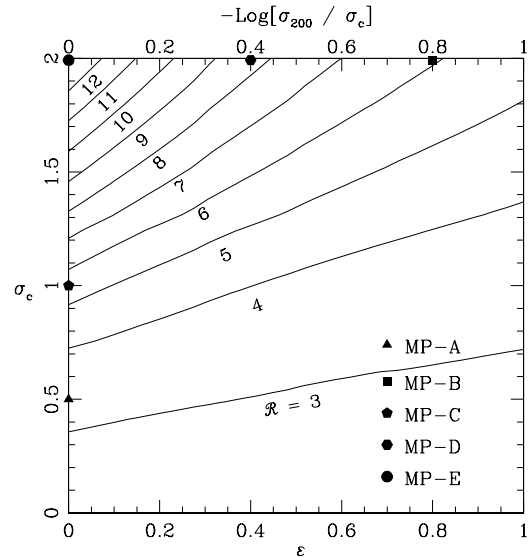} }
\vskip -0.2 truecm
\caption{
Contours of constant hardness ratio ${\cal R}$, equation~(\ref{hardness}). 
}
\label{Rplot}
\end{figure}
 
Given the very different behavior of the low and high energy line
emission, we investigate the behavior of a hardness ratio 
\beq
{\cal R} \ = \ \frac{L_{tot}[0.6-1.5 \kev]}{L_{tot}[6.6-7.5 \kev]}
\label{hardness}
\eeq
in the multiphase model plane.  For the single phase case, ${\cal R} =
2.69$.  
Contours of constant ${\cal R}$ in the $\{ \sigma_c, \epsilon \}$ 
plane are shown in Figure~\ref{Rplot}.  From this figure, it is
clear that even moderate signal-to-noise spectra could produce useful
constraints among models with similar baryon fraction correction
factors.  Models A and B differ in ${\cal R}$ by nearly a factor 2.
Models B and C are nearly degenerate in this measure, but inspection
of Figure~\ref{rsband} shows that model C is more continuum dominated
at low energies while model B has a larger line contribution.

\subsection{Limiting the baryon fraction bias }

The concluding sentence of GT96 expresses a view ``that the 
intracluster medium is much more complex than most people have
hitherto assumed and that there is sufficient uncertainty in its
modeling to permit a critical density, Einstein-deSitter universe''.  
Within the context of the expanded version of their model which we
develop here, we can ask whether this opinion is supported by recent 
data.  

A number of high quality measurements of the Hubble constant 
from SZ and \xray observations have been made recently (Myers \etal
1997; Birkinshaw \& Hughes 1994; Jones 1995; Hughes \& Birkinshaw
1998; Holzapfel et al. 1997).  Hughes \& Birkinshaw (1998) present an
ensemble value $\Ho \se 47 \pm 7 \kmsmpc$ from these studies.  
For a true value $\Ho \se 65 \kmsmpc$, supported by Type Ia supernovae 
(Hamuy \etal 1995; Riess \etal 1996), expanding photosphere of Type
II supernovae (Schmidt \etal 1994) gravitational lens time delays 
(Kundi \etal 1997; Schechter \etal 1997), this ensemble average is low
by 28\%.  Assuming that a multiphase structure is at least 
partly responsible for this biased estimate --- other effects may lead to
an underestimate at the $\sims 5-10\%$ level (Cen 1998) ---  
then a bound on the SZ decrement enhancement 
$\Delta y/y_s \lta 0.14$ results.  Comparing the contours in
Figures~\ref{cbarycont} and \ref{SZcontor}, this limit 
restricts baryon fraction diminution factors to be modest, 
$0.75 \lta C_{\rm b} \le 1$.  

\subsection{Caveats and extensions }

The model we present contains a number of simplifying assumptions.  It
is important to note that the model is, in principle, falsifiable.
Given a known Hubble constant, likelihood analyses of SZ observations 
and \xray spectra will independently identify preferred regions in 
the $\{ \sigma_c, \epsilon \}$ plane.  If these regions are
consistent, the observations can be combined to yield a best estimate
location $\{ \hat{\sigma}_c, \hat{\epsilon} \}$, and an estimate of 
the baryon fraction bias $C_{\rm b}$ (Figure~\ref{cbarycont}).
Inconsistent constraints may imply a need to relax one or more of the
following assumptions.  

\setlength{\parskip}{0.10in}
\noindent
{\sl Lack of spherical symmetry.\/} 
With rare exceptions, cluster \xray images are close to round.  Most 
have axial ratios $b/a \gta 0.8$ (Mohr \etal 1995).  Such small deviations
from spherical symmetry lead to scatter, but little bias, in
determinations of $\Ho$ from SZ+\xray analysis (Sulkanen, 1999).  
Given supporting evidence for a multiphase ICM in a cluster of
moderate ellipticity, the spherical model introduced here could be
extended to prolate or oblate spheroids.  A more profitable
approach might be to include multiphase structure in the deprojection
method discussed by Zaroubi \etal(1998). 

\noindent
{\sl Non-isothermal emission weighted temperature profiles.\/} 
There is indication from ASCA observations (Markevitch \etal 1998)
that the emission weighted ICM temperature declines substantially
within the virial radius.  However, ROSAT colors rule out a
temperature drops of $12/20\%$ for $5/10 \kev$ clusters within
one-third of $r_{200}$ (Irwin, Bregman \& Evrard 1999).  It is
straightforward to include a radial temperature gradient $T_X(r)$ 
into the analysis, entering into the definition of the 
pressure profile, equation~\ref{pdef}.

\noindent
{\sl Non-lognormal distribution of density fluctuations.\/} 
The chosen form of the density distribution is motivated by simplicity
and by the observation that non--linear gravity on a Gaussian random density 
field characteristically generates a log-normal pdf (Cole \& Weinberg
1994).  The results are sensitive to low order moments of the
density distribution.  We await observations and future numerical 
simulations including cooling and galaxy-gas interactions in a 
three-dimensional setting to shed light on the appropriate form of the
density fluctuation spectrum. 

\noindent
{\sl Non-isobaric equation of state.\/}
This may be the most readily broken of our model assumptions.  The
cluster environment is very dynamic.  During large mergers, the behavior of
the gas in the inner regions of infalling subclusters is essentially 
adiabatic (Evrard 1990; Navarro, Frenk \& White 1993).  During
quiescent periods between mergers, a cluster atmosphere may stabilize
and develop the assumed isobaric perturbations during a cooling flow
phase (Thomas \etal 1986).  In the transition period, isobaric
perturbations in the core may co-exist with adiabatic perturbations
near the virial radius.  Empirical constraints will come from
improved spectroscopic imaging. 

\noindent
{\sl Binding mass estimates under hydrostatic equilibrium.\/} 
As noted in \S2.2, the radial dependence of our model multiphase 
distribution can lead to a total mass profile $M_{tot,m} (r)$ 
that is steeper than that determined for a single-phase gas. 
Galaxy cluster lensing observations could be used to test the mass
distribution predicted by multiphase models (see Figure \ref{totmass}).
This provides another independent constraint on the admissable
region of the multiphase $\{\sigma_c,\epsilon\}$ plane.
\setlength{\parskip}{0.0in}

\section{Summary and Discussion}

We present a spherically symmetric, multiphase model of the
intracluster medium in galaxy clusters.  The model assumes existence
of a lognormal distribution of isobaric density and temperature 
fluctuations at any radius.  The radially dependent 
variance of the density fluctuations $\sigma^2(r)$ is subject to 
two empirical constraints : 1) that the broadband \xray emissivity
profile matches observations and 2) that the \xray emission-weighted 
temperature is constant with radius.  

We calculate the bias introduced in cluster gas mass fraction
estimates when a single-phase model is a applied to a multiphase
atmosphere.  As derived by GT96, the standard
analysis of the \xray observations with a single-phase assumption will 
overestimate the baryon fraction in the multiphase case.   
Examining observable effects on the 
central Sunyaev-Zel'dovich  decrement as well as \xray spectroscopy, 
we demonstrate how, within the context of this model, 
the bias can be recovered by existing and future observations.  

Large values of the clumping factor ${\cal C}$, hence 
large reduction in the cluster baryon fraction are not favored by
current observations.  Models with high values of $\sigma_c$ 
produce a nearly power-law \xray bremsstrahlung 
continuum and bias estimates of the Hubble constant.  An ensemble
mean value $47 \pm 7 \kmsmpc$ (Hughes \& Birkinshaw 1998) 
arising from recent SZ+\xray analysis,
when compared to an assumed value of $65 \kmsmpc$, 
suggests clumping only overestimates ICM gas fractions by $\lta 20\%$.  

Spatially resolved \xray spectroscopy, particularly of line 
emission in cooler region (0.1-3keV), will provide tests of multiphase 
model.  Data from the upcoming AXAF and XMM missions will be
particularly valuable.

\section*{Acknowledgments}
This work is supported by the National Science Foundation through the 1998 
physics REU summer program at the University of Michigan and through
grant AST-9803199.  We acknowledge NASA support through grant 
NAG5-7108 and NSF through grant AST-9803199.  M.E.S. thanks
NASA's Interagency Placement Program, the University of Michigan Department
of Astronomy, and the University of Michigan Rackham Visiting Scholars 
Program.

\end{document}